\documentclass[letterpaper]{sig-alternate}

\usepackage{ifpdf}
\ifpdf
\else
\fi

\usepackage{graphicx}
\usepackage{url}
\usepackage[colorlinks, citecolor=black, urlcolor=black, linkcolor=black]{hyperref}
\usepackage[small]{caption}

\begin{document}

\conferenceinfo{IATP 2013, }{August 1, 2013, Dublin, Ireland.}
\CopyrightYear{2013}

\title{Adaptive Keywords Extraction with Contextual Bandits for Advertising on Parked Domains}

\numberofauthors{2} \author{
\alignauthor
Shuai Yuan, Jun Wang\\
       \affaddr{Department of Computer Science, \\University College London}\\
       \affaddr{London, United Kingdom}\\
       \email{\{s.yuan, j.wang\}@cs.ucl.ac.uk}
\alignauthor
Maurice van der Meer\\
       \affaddr{B.V. DOT TK}\\
       \affaddr{Amsterdam, Netherlands}\\
       \email{maurice@dot.tk}
}


\maketitle

\begin{abstract}

  Domain name registrars and URL shortener service providers place advertisements on the parked domains (Internet domain names which are not in service) in order to generate profits. As the web contents have been removed, it is critical to make sure the displayed ads are directly related to the intents of the visitors who have been directed to the parked domains. Because of the missing contents in these domains, it is non-trivial to generate the keywords to describe the previous contents and therefore the users intents. In this paper we discuss the adaptive keywords extraction problem and introduce an algorithm based on the BM25F term weighting and linear multi-armed bandits. We built a prototype over a production domain registration system and evaluated it using crowdsourcing in multiple iterations. The prototype is compared with other popular methods and is shown to be more effective.


\end{abstract}

\section{Introduction}\label{section-introduction}

\begin{figure}[t]
	\centering
	\includegraphics[width=.4 \textwidth]{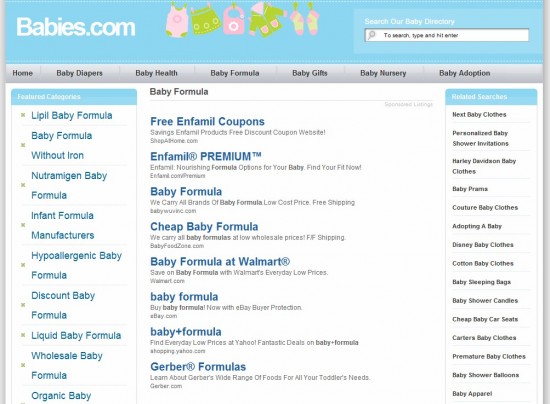}
	\caption{An example of advertising on parked domains. Instead of showing an error or \textit{under construction}, relevant ads are displayed.}
	\label{fig-parked_domain_example}
\end{figure}

Online advertising is a fastest growing area in IT industry in recent years. In an online advertising eco-system, yield optimisation is the core problem of the supply side, aka, publishers, and is commonly dealt with by managing revenue channels \cite{roels2009dynamic}; selecting relevant ads \cite{yuan2012sequential}; optimising reserve prices \cite{ostrovsky2009reserve}; controlling advertising level \cite{dewan2003management}, etc. The work presented in this paper is focused on a specific problem: advertising on parked domains. Quite often, domains hold valid web contents until certain stages when they are bought/revoked/suspended/expired later. Before those domains are (re)used, they are empty with no content displayed but ads. This occurs commonly in the URL shortener services and domain registration services. As people may not know the webpages do not exist any more, there are still many in-links available online which generate good amount of traffic towards the parked domains. As a general practice, the domain providers or registrars host display ads on these parked domains in order to generate profit, c.f. Figure~\ref{fig-parked_domain_example}.  Ideally the ads are required to be as relevant as possible to the visitors. A clear difference from the existing contextual advertising research \cite{yih2006finding, broder2007semantic} is that the context (webpage content) is no longer available when impressions are created and therefore it is hard to generate keywords to describe potential visitors (interests or intents).

In this paper, we use contextual multi-armed bandits \cite{langford2007epoch, li2010contextual} to predict the relevant keywords over time.  Different from the standard multi-armed bandits, every arm is associated with a feature vector as the side information known to the player. We employ the general linear model (GLM) to describe the relationship of rewards (relevance) and features (keywords), where the model may also be referred to as the linear bandits \cite{filippi2010parametric}. We take the upper confidence bound (UCB) \cite{auer2003using} approach to explore and exploit optimal arms.
Moreover, we applied crowdsourcing to collect user feedback and adjust the algorithm. 
We chose accessors randomly from \url{oDesk.com}, one of the largest online crowdsourcing services, given they had the sufficient knowledge to understand the webpages and their tasks. 
Besides screenshots and remote desktop monitoring, we embedded multiple traps in judge items and considered the user failed the task if 30\% traps were trigged. 
To validate the results, we employed both Fleiss' Kappa \cite{joseph1971measuring} and Cohen's Kappa \cite{cohen1960coefficient} to measure the agreement of judgements. Our experiments show the effectiveness of the proposed approach in addressing the problem.


\section{Related works}

There are commercial tools generating high co-occurrence or similar phrases as suggestions based on seed terms provided by user, like Google AdWords Keyword Tool\footnote{\url{adwords.google.com/select/KeywordToolExternal}}. There are two problems for these tools: 1) it still requires fair amount of manual work to input seeds and choose from suggestions. 
2) the topics of generated phrases are based on user query logs, existing bid phrases, and lexical analysis, and may easily drift from the original from webpages.

Keywords extraction is largely considered a supervised learning problem \cite{turney2000learning, hulth2003improved, wu2009advertising, ravi2010automatic} where the algorithms learn to classify as positive or negative examples of keywords based on training sets. These algorithms need expensive human labelled dataset in advance, and usually perform the learning process offline. The model could become inaccurate when users' interests change over time. In \cite{hulth2003improved} linguistic knowledge is introduced such as noun-phrase-chunks (NP-chunks) and part-of-speech (POS) to outperform using statistics features only. These linguistic features are also used in our work. Besides, query logs are also used as a good reflect of users' interests \cite{fuxman2008using}.

There are also extensive research about keywords suggestion. In \cite{chen2008advertising} the keywords are suggested based on concept hierarchy mapping therefore the suggestions are not limited to the bag-of-words of the webpage, and may expand to non-obvious ones which are categorized in bigger concepts. The work of \cite{joshi2006keyword, abhishek2007keyword} recognise that the bid prices for hot keywords are high therefore would cost more, and try to find related non-obvious keywords that are cheaper. Although these keywords may have lower traffic, but when combined the traffic could match that of a hot one, while these keywords cost still less.

The work of \cite{yih2006finding} 
proposed a classifier that uses multiple text features, including how often the term occurs in search query logs, to extract keywords for ads targeting based on logistic regression. The system discussed in \cite{ravi2010automatic} first generates candidates by several methods including a translation model capable of generating phrases not appearing in the text of the pages. Then candidates are ranked in a probabilistic framework using both the translation model favouring relevant phrases, as well as a language model favouring well-formed phrases. Another relevant work can be found in \cite{rusmevichientong2006adaptive}. The authors proposed an exploration-exploitation algorithm of sorting keywords in an descending order of profit-to-cost ratio and adaptively identify the set of keywords to bid on based on historical performance, with a daily budget constraint. In this paper we try to find the keywords from the given webpage with additional web knowledge, rather than find profitable ones from a very large set (like 50k). Then we leave the matching between keywords and ads to display networks/exchanges, such as Google AdSense.

In \cite{li2010exploitation} the authors propose a combination algorithm of using upper confidence bound (UCB) and $\epsilon$-greedy to solve the exploration-exploitation dilemma. Their target is to select high profit ads which would be fed in contextual advertising platforms. The feature vectors of ads are not used in their system as side information, instead they use standard bandits considering the reward following an unknown stochastic process. Our research is partially inspired by their work.

\section{Online Keywords Extraction}
\label{section-problem}

\begin{figure}[t]
	\centering
	\includegraphics[width=.5 \textwidth]{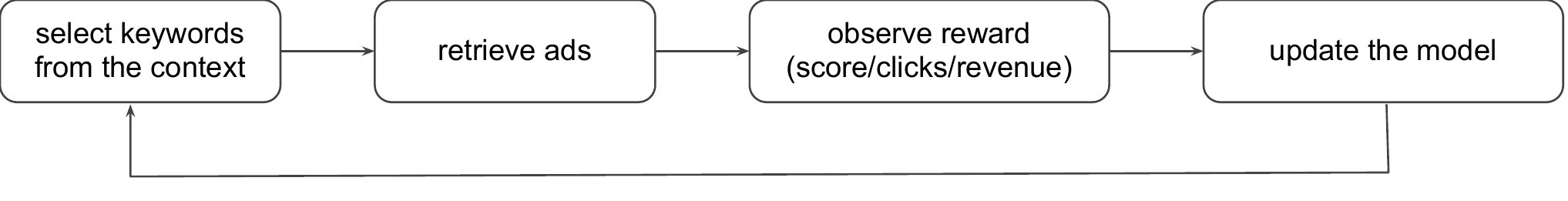}
	\caption{The keywords extraction task could be an iterative process. The publisher would like to exploit current optimal keywords as well as to explore potentially better ones.}
	\label{fig-flowchart}
	\vspace{-15px}
\end{figure}

From the publishers' perspective the process of extracting keywords is illustrated in Figure-\ref{fig-flowchart}. It is an iterative process of selecting candidates and update the model according to the feedback. This task of extracting keywords against user feedback is directly linked to the contextual multi-armed bandits problem. The linear multi-armed bandits used in our work \cite{dani2008stochastic, rusmevichientong2010linearly, abbasi2009forced, auer2003using} is a special case of contextual bandits. It has the same setting with standard bandits except the availability of the side information. The side information of each arm determines the reward for pulling the arm therefore could be used to make decisions. In our work, we mainly exploited text features from webpages under domains being evaluated, including TF-IDF scores in title, content, keyword and description in HTML meta tag, header, anchor text of in-links, and part of speech. Since we hold the webpages or redirections of parked domains, extracting these features is possible.

First of all we define reward of iteration step $t \in [0, T]$ as a real number in [0,1],
\begin{equation}
r(t) \in [0,1]
\end{equation}

The reward could be in various form, for example 1) relevance scores of selected keywords against the webpage; 2) the clickthrough rates (CTR) of ads, or 3) profit gained in each iteration. Apparently the relevance scores, CTRs and profit are linked, however not guaranteed with any kind of linear/non-linear relationship to our best extent of knowledge. 
In our work we used human judged relevance score for simplicity, also due to the difficulty of acquiring necessary data of CTR and profit. Besides, we employed the crowdsourcing approach to get the feedback in a cheap and fast way. We considered each webpage a K-armed bandit machine and every arm a word (or a multi-word phrase). In each iteration an arm will be pulled resulting in selection of the corresponding word/phrase to be the keywords of the webpage. Different from standard bandits, each arm ($i \in K$) of the linear bandits is associated with some $D$-dimension feature vector $\mathbf{x_i} \in \mathbb{R}^{D\times 1} $ which is already known. The expected reward (user feedback) is given by the inner product of its feature vector $\mathbf{x_i}$ and some fixed, but initially unknown parameter (column) vector $\mathbf{w}$. That is, the reward is a linear function of the feature vector and unknown parameters, 
\begin{equation}
\label{equation-linear} 
r_i = \mathbf{x_i}' \cdot \mathbf{w} 
\end{equation}

The goal here is to get the optimal reward. We take the \textsc{LinRel} algorithm \cite{auer2003using} highlighting upper confidence bound (UCB) to deal with the exploration-exploitation dilemma.

\subsection{Upper Confidence Bound Approach}

The idea of \textsc{LinRel} algorithm is to estimate the reward for the $i$-th arm from the linear combination of historical reward received, with a high probability. We expand the \textsc{LinRel} algorithm a bit to help the paper self-contained. First we write the feature vector of the $i$-th arm as the linear combination of the previous chosen feature vectors (this is always possible except for the initial $d$ iteration steps),
\begin{equation}
\mathbf x_i(t) = \mathbf X(t) \cdot \mathbf a_i(t)
\end{equation}
where $\mathbf a_i(t) \in \mathbb{R}^{t \times 1}$ is the coefficient of the linear combination and $\mathbf X(t)$ is the feature matrix selected in past iteration steps with dimension $D\times t$,
\begin{equation}\label{equation-X}
\mathbf X(t) = [\mathbf x(1), \mathbf x(2), \ldots, \mathbf x(t-1)]
\end{equation}
where the $\mathbf x(t)$ denotes the feature vector used in $t$ iteration step (without specifying the selected arm). With the same coefficient the reward could be written as,
\begin{equation}
r_i(t) = \mathbf x_i(t)' \cdot \mathbf w = (\mathbf X(t) \cdot \mathbf a_i(t))' \cdot \mathbf w = \mathbf R(t)' \cdot \mathbf a_i(t)
\end{equation}
where $\mathbf R(t) \in \mathbb{R}^{t \times 1}$ is a vector of historical reward,
\begin{equation}\label{equation-R}
\mathbf R(t) = [r(1), r(2), \ldots, r(t-1)]'
\end{equation}
This gives a good estimate $\mathbf R(t) \cdot \mathbf a_i(t)$ for $r_i(t)$. The algorithm keeps the variance small to maintain a narrow confidence interval of the estimate. By assuming i.i.d of $r_i(t)$ (which is true for our keywords extraction task) and since $r_i(t) \in [0,1]$ the variance of this estimate is bounded by $\|\mathbf a_i(t)\|^2/4$.
In order to get $\mathbf a_i(t)$ first calculate the eigenvalue decomposition,
\begin{equation}\label{equation-eigendecomposition}
\mathbf X(t) \cdot \mathbf X(t)' = \mathbf U(t)' \cdot \mathbf \Delta[\lambda_1, \lambda_2, \ldots, \lambda_d] \cdot \mathbf U(t)
\end{equation} 
where $\lambda_1, \ldots, \lambda_k \geq 1$ and $\lambda_{k+1}, \ldots, \lambda_d <1$. Then for each feature vector $\mathbf x_i(t)$ write,
\begin{align}
\mathbf z_i(t) &= \mathbf U(t) \cdot \mathbf x_i(t) = [x_{i,1}(t), x_{i,2}(t), \ldots, x_{i, d}(t)]' \\
\mathbf u_i(t) &= [x_{i,1}(t), \ldots, x_{i, k}(t), 0, \ldots]' \\
\mathbf v_i(t) &= [0, \ldots, x_{i,k+1}(t), \ldots, x_{i, d}(t)]'
\end{align}
The coefficient $\mathbf a_i(t)$ is calculated as,
\begin{equation}\label{equation-a}
\mathbf a_i(t) = \mathbf u_i(t)' \cdot \mathbf \Delta[\frac{1}{\lambda_1}, \cdots, \frac{1}{\lambda_k}, \cdots, 0] \cdot \mathbf U(t) \cdot \mathbf X(t)
\end{equation}
Then with probability of $1-\delta/T$ the expected reward of $i$-th arm is,
\begin{equation}\label{equation-ucb}
ucb_i(t) = \mathbf R(t)\cdot \mathbf a_i(t) + \sigma
\end{equation}
where $\sigma$ is the width of the confidence bounds by using Azuma-Hoeffding bound \cite{azuma1967weighted},
\begin{equation}\label{equation-sigma}
\sigma = \|a_i(t)\|(\sqrt{ln(2TK/\delta})+\|v_i(t)\|
\end{equation}
where $\delta$ is used to confine a narrow confidence interval and recall $K$ is the number of arms. The arm with highest $ucb_i(t)$ score will be selected. Apparently the arm got selected because the combination of its expected reward ($\mathbf R(t)\cdot \mathbf a_i(t)$) or potential reward ($\sigma$) is high. The former leads to exploitation and the latter results in exploration.

\section{Experiments and Results}

In this section we introduce the architecture of the prototype system and its evaluation against BM25F \cite{zaragoza2004microsoft}, KEA \cite{witten1999KEA}, and two popular web services.

\subsection{Prototype System}

\begin{figure}[t]
	\centering
	\includegraphics[width=.35 \textwidth]{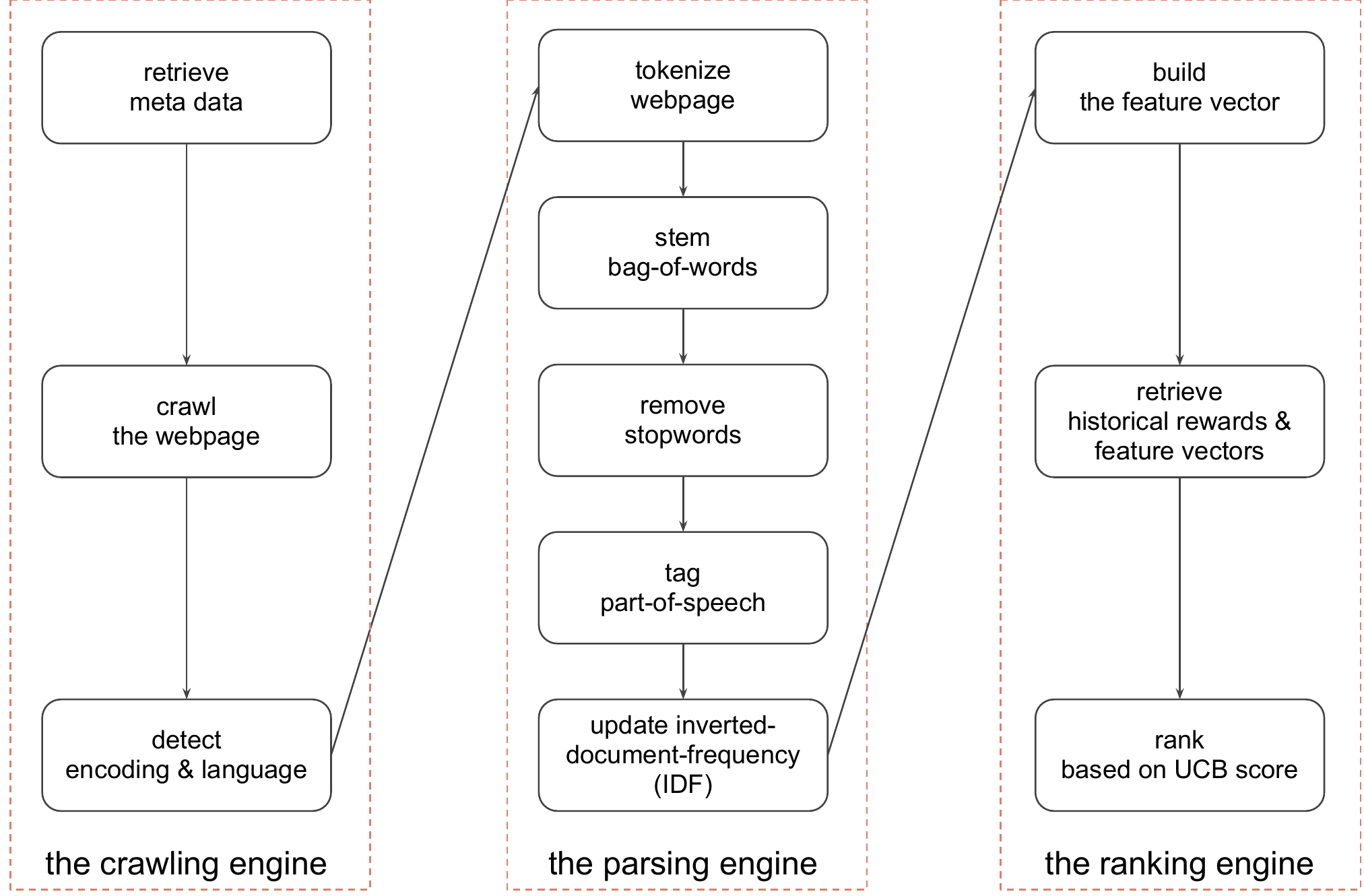}
	\caption{The architecture of the \textsc{razorclaw} prototype system. We divide the system into 3 main parts. In every part there are loosely bounded modules with each for a single task. These modules could be changed or updated flexibly. We intent to create an open framework for the keywords extraction task.}
	\label{fig-system-architecture}
	\vspace{-5px}		
\end{figure}

We have built a prototype system \textsc{razorclaw} according to the model discussed above. The system is open source, written in Java, and online running. Its architecture is illustrated in Figure-\ref{fig-system-architecture}. Generally the prototype system is made of 3 main parts: the crawler, parser, and the ranker. In the parsing engine, different modules will be used for different languages.

In the crawling module, the \textsc{razorclaw} first loads meta information from the parked domain database, including the previously displayed webpage, the anchor texts and referrer URLs collected from the Internet. The webpage is crawled, too. The next step is to detect the correct encoding of the text (sometimes not presented in HTML) and the language. The language detector we use\footnote{\url{code.google.com/p/language-detection}} is based on naive Bayesian filter and has reported 99\% precision over 49 languages. 

Once the encoding and language are detected we convert the content to UTF-8 and invoke corresponding NLP processor. We employ openNLP\footnote{\url{incubator.apache.org/opennlp}} for major western languages and IKAnalyzer\footnote{\url{code.google.com/p/ik-analyzer}} for Chinese-Japanese-Korean-Vietnam languages. The language support is not the main point of the paper however our system processes more than 50 languages which is especially useful for free domain services, where we discovered a great portion of registrations are from south east Asia, India, and Arabic countries. 

After stemming, removing stopwords, part-of-speech (POS) tagging and saving the inverted-document-frequency (IDF), each phrase in the bag-of-words is processed to generate its local feature vector and to retrieve historical features and rewards. According to the ranking result our system gives candidate keywords (usually the top 3).

\begin{figure}[t]
	\centering
	\includegraphics[width=.35 \textwidth]{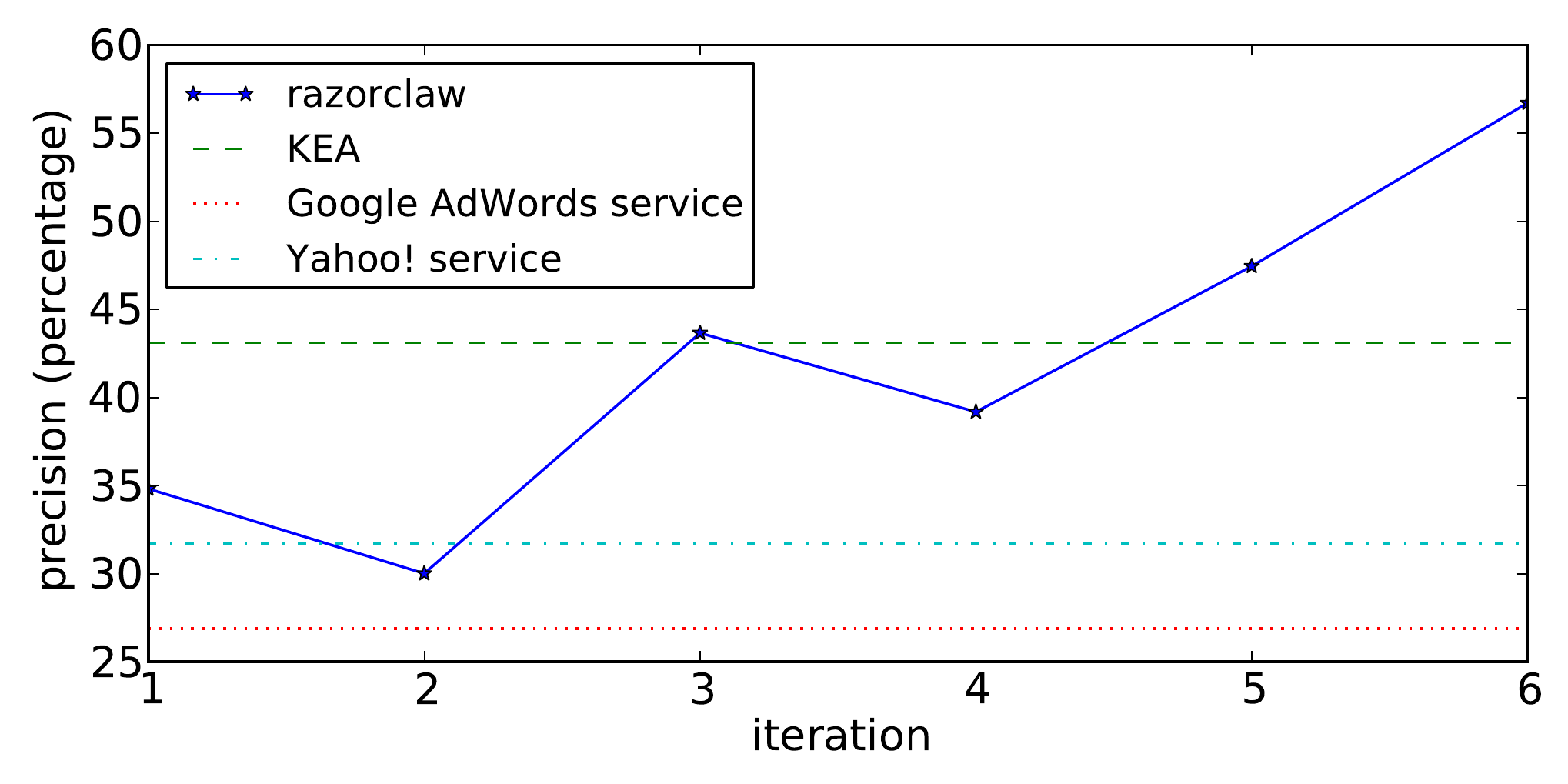}
	\vspace{-10px}
	\caption{The performance comparison of competing algorithms and services.}
	\vspace{-15px}	
	\label{fig-precision}
\end{figure}

\begin{figure}[t]
	\centering
	\includegraphics[width=.4 \textwidth]{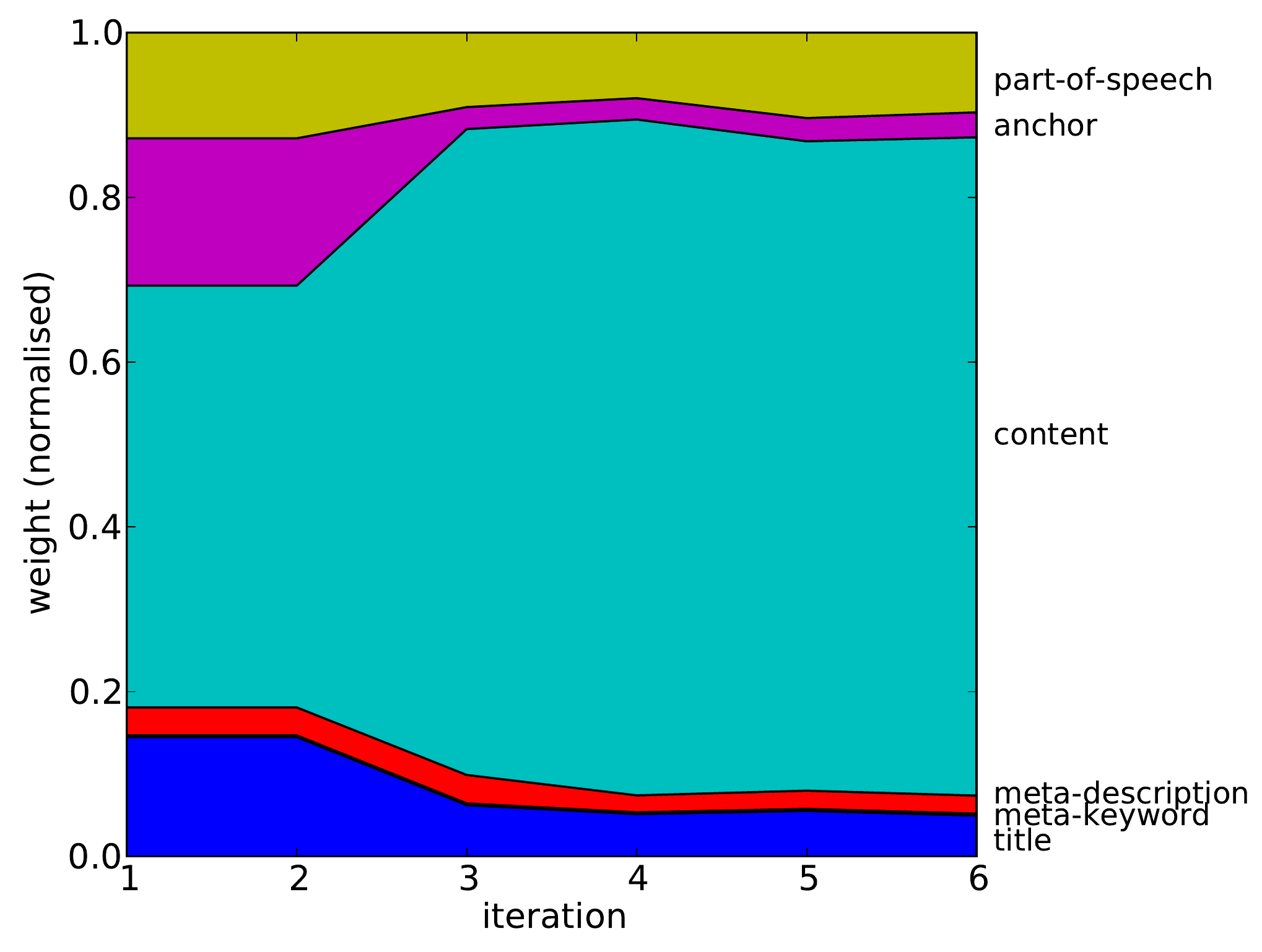}
	\vspace{-10px}
	\caption{The evolution of weights of some features of razorclaw system. Features with too small weights are not included here. The initial weights were obtained from \cite{zaragoza2004microsoft}. The result showed that the \textit{content} of webpage is the most influential factor, followed by \textit{part-of-speech} and \textit{title}. The \textit{keyword} and \textit{description} in the HTML meta tag are in fact not trustworthy.}
	\vspace{-5px}	
	\label{fig-weight}
\end{figure}

\begin{figure}[t]
	\centering
	\includegraphics[width=.35 \textwidth]{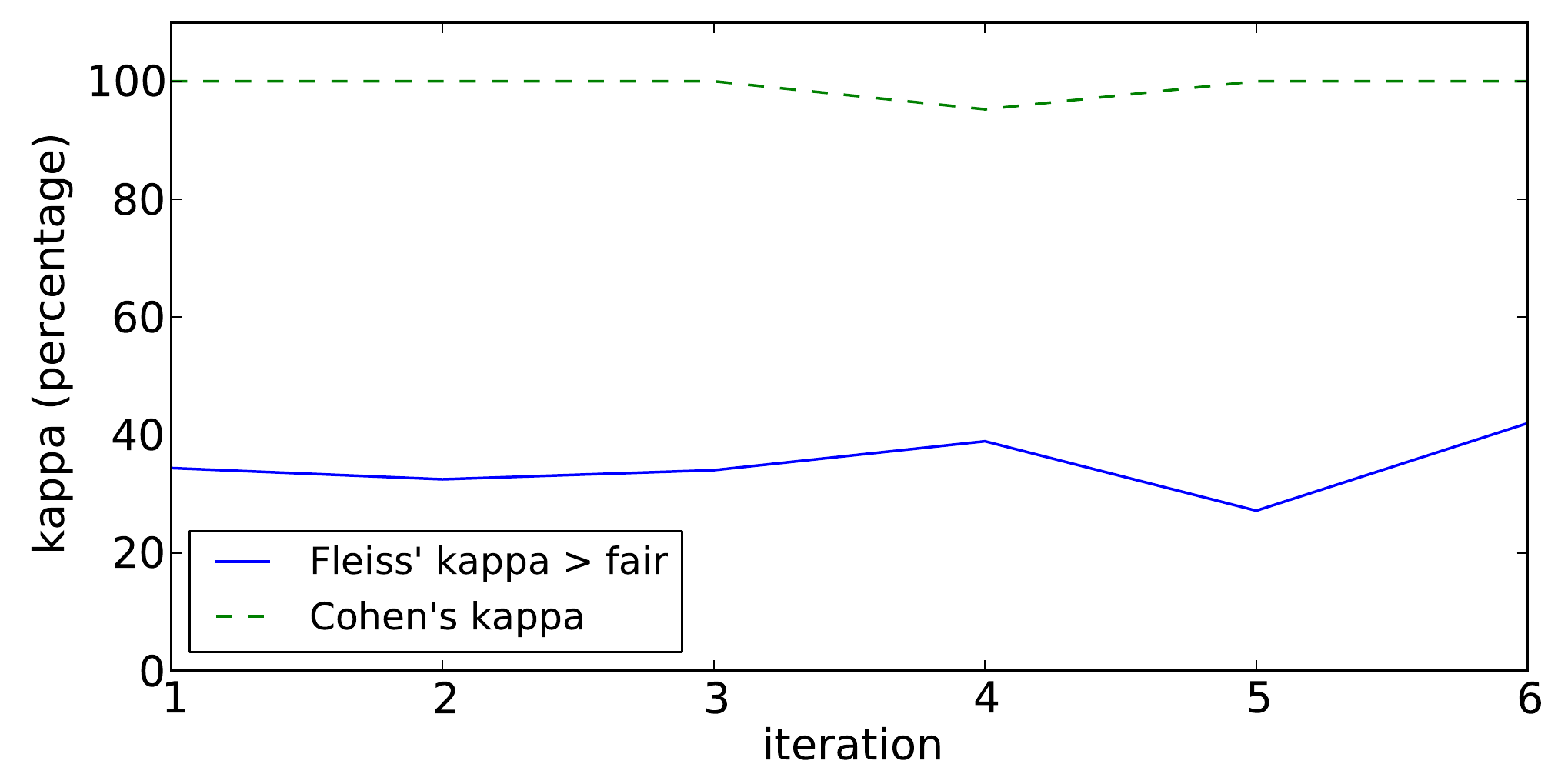}
	\vspace{-10px}
	\caption{The plot of agreement measurement for crowdsourcing ranking results. The \textit{fair} in Fleiss' kappa corresponds to 0.2}
	\vspace{-15px}		
	\label{fig-kappa}
\end{figure}

\subsection{Datasets and Algorithms}

We used 165 domains from Dot.tk's database and collected all possible side information. We compared our prototype system with BM25F algorithm \cite{zaragoza2004microsoft}, KEA algorithm\footnote{\url{www.nzdl.org/kea}} \cite{witten1999KEA}, Yahoo! Term Extraction service\footnote{\url{goo.gl/KXaPw}} and Google Targeting Idea service\footnote{\url{goo.gl/CpNgp}}. 

The parameters of BM25F algorithm was obtained from \cite{zaragoza2004microsoft}. KEA is a famous keywords extraction tool in academic research. It is capable of extraction keywords with domain specific knowledge, for example with Wikipedia article names. However in our experiments we select domains randomly and do not restrict the result to any specific area. 

Both web services require registration as a developer and incur cost if exceeding the free quota. 
According to the specification, we send text content of webpages (including the title and meta) to the Yahoo! service and send URLs to the Google one.

\subsection{Crowdsourcing Based Experiments}

In order to carry out the evaluation in an efficient and cheap fashion, we employed a crowdsoucing platform to judge the results of multiple algorithms. We conducted a test on the popular crowdsourcing platform oDesk.com and randomly selected 23 human assessors who passed the test. We then asked them to rank the relevance between the domain and keywords on a 1 (least relevance) to 5 (most relevance) scale. We gave very detailed instructions to help keep the ranking as consistent as possible among different assessors and in different iteration steps. For broken links or empty keywords (some algorithm failed to give the result) the assessors were asked to give a 0.  We chose oDesk.com for its simplicity, openness, big candidate pool, and various tools to monitor the progress. We designed traps, used screenshots and remote desktop monitoring to reduce fraud and malicious behaviours. Among the returns 3 users were marked careless (by trigging 3 out of 10 traps) and their results were completely removed from the pool.

\subsection{Results}

We carried out 6 iterations of experiments in total, on 7 June, 17 July, 4 August, 11 August, 15 August, and 20 August 2011. The comparison of competing algorithms is reported in Figure~\ref{fig-precision}. As introduced above, the relevance ranking ranges from 0-5 with 0 is specifically for failures (website did not open, encoding or language did not recognised, or other issues resulting in failure of extracting keywords). In every iteration each domain was ranked by at least 2 assessors.

First we compare the overall precision (cumulative relevance ranking against the highest possible score). From Figure~\ref{fig-precision} we can see the \textsc{razorclaw} performed significantly better than the 2nd best (KEA), with 8.29\% behind in the first iteration, but 13.6\% improvement in the last one. The web services performed less satisfying mainly due to they were limited to English language at the time of test.

From the precision plot the improvement of \textsc{razorclaw} is clearly observable. At iteration 2 and 4 the algorithm performed worse than the previous round, suffering from the exploration. However, the algorithm was able to pick up new sets of weights of features as anticipated. From Figure~\ref{fig-weight} we can see the corresponding evolution of the weights. In the figure we listed top 6 features and did not include ones with too small weights. The \textit{content} was the most influential factor, followed by \textit{part-of-speech} and \textit{title}. The \textit{keyword} from the HTML meta tag is in fact not trustworthy, which is consistent with the no-use decision in major search engines like Google \cite{meta-google}. The measurement of agreement is plotted in Figure~\ref{fig-kappa}. We computed the Cohen's kappa of any two assessors and reported its average of each iteration. Similarly the Fleiss' kappa of all assessors of each iteration was plotted. Generally speaking the agreement of results was satisfying.

\section{Conclusion}
In this paper, we discussed an adaptive keyword extraction algorithm based on BM25F and contextual multi-armed bandits. We showed its good performance in crowdsourcing based experiments, and reported interesting findings including the evolution of weights. We plan to include more features and conduct larger scale of evaluation in future.

{\tiny
\bibliographystyle{abbrv}
\bibliography{bib}

\begin{thebibliography}{10}
\vspace{15px}

\bibitem{meta-google}
Meta tags.
\newblock \url{goo.gl/iTgs9} (last visited 20/6/2013).

\bibitem{abbasi2009forced}
Y.~Abbasi-Yadkori, A.~Antos, and C.~Szepesv{\'a}ri.
\newblock Forced-exploration based algorithms for playing in stochastic linear
  bandits.
\newblock In {\em Proceedings of the COLT 2009}.

\bibitem{abhishek2007keyword}
V.~Abhishek and K.~Hosanagar.
\newblock Keyword generation for search engine advertising using semantic
  similarity between terms.
\newblock In {\em Proceedings of the ACM EC 2007}.

\bibitem{auer2003using}
P.~Auer.
\newblock Using confidence bounds for exploitation-exploration trade-offs.
\newblock {\em The Journal of Machine Learning Research}, 2003.

\bibitem{azuma1967weighted}
K.~Azuma.
\newblock Weighted sums of certain dependent random variables.
\newblock {\em Tohoku Mathematical Journal}, 1967.

\bibitem{broder2007semantic}
A.~Broder, M.~Fontoura, V.~Josifovski, and L.~Riedel.
\newblock A semantic approach to contextual advertising.
\newblock In {\em Proceedings of the ACM SIGIR 2007}.

\bibitem{chen2008advertising}
Y.~Chen, G.-R. Xue, and Y.~Yu.
\newblock Advertising keyword suggestion based on concept hierarchy.
\newblock In {\em Proceedings of the ACM WSDM 2008}.

\bibitem{cohen1960coefficient}
J.~Cohen et~al.
\newblock A coefficient of agreement for nominal scales.
\newblock {\em Educational and psychological measurement}, 1960.

\bibitem{dani2008stochastic}
V.~Dani, T.~P. Hayes, and S.~M. Kakade.
\newblock Stochastic linear optimization under bandit feedback.
\newblock In {\em Proceedings of the COLT 2008}.

\bibitem{dewan2003management}
R.~M. Dewan, M.~L. Freimer, and J.~Zhang.
\newblock Management and valuation of advertisement-supported web sites.
\newblock {\em Journal of Management Information Systems}, 2003.

\bibitem{filippi2010parametric}
S.~Filippi, O.~Capp{\'e}, A.~Garivier, and C.~Szepesv{\'a}ri.
\newblock Parametric bandits: The generalized linear case.
\newblock {\em Advances in Neural Information Processing Systems}, 2010.

\bibitem{fuxman2008using}
A.~Fuxman, P.~Tsaparas, K.~Achan, and R.~Agrawal.
\newblock Using the wisdom of the crowds for keyword generation.
\newblock In {\em Proceeding of the ACM WWW 2008}.

\bibitem{hulth2003improved}
A.~Hulth.
\newblock Improved automatic keyword extraction given more linguistic
  knowledge.
\newblock In {\em Proceedings of the EMNLP 2003}.

\bibitem{joseph1971measuring}
F.~L. Joseph.
\newblock Measuring nominal scale agreement among many raters.
\newblock {\em Psychological bulletin}, 1971.

\bibitem{joshi2006keyword}
A.~Joshi and R.~Motwani.
\newblock Keyword generation for search engine advertising.
\newblock In {\em Proceedings of the ICDM 2006}.

\bibitem{langford2007epoch}
J.~Langford and T.~Zhang.
\newblock The epoch-greedy algorithm for contextual multi-armed bandits.
\newblock {\em Advances in Neural Information Processing Systems}, 2007.

\bibitem{li2010contextual}
L.~Li, W.~Chu, J.~Langford, and R.~E. Schapire.
\newblock A contextual-bandit approach to personalized news article
  recommendation.
\newblock In {\em Proceedings of the ACM WWW 2010}.

\bibitem{li2010exploitation}
W.~Li, X.~Wang, R.~Zhang, Y.~Cui, J.~Mao, and R.~Jin.
\newblock Exploitation and exploration in a performance based contextual
  advertising system.
\newblock In {\em Proceedings of the ACM SIGKDD 2010}.

\bibitem{ostrovsky2009reserve}
M.~Ostrovsky and M.~Schwarz.
\newblock Reserve prices in internet advertising auctions: A field experiment.
\newblock 2009.

\bibitem{ravi2010automatic}
S.~Ravi, A.~Broder, E.~Gabrilovich, V.~Josifovski, S.~Pandey, and B.~Pang.
\newblock Automatic generation of bid phrases for online advertising.
\newblock In {\em Proceedings of the ACM WSDM 2010}.

\bibitem{roels2009dynamic}
G.~Roels and K.~Fridgeirsdottir.
\newblock Dynamic revenue management for online display advertising.
\newblock {\em Journal of Revenue \& Pricing Management}, 2009.

\bibitem{rusmevichientong2010linearly}
P.~Rusmevichientong and J.~N. Tsitsiklis.
\newblock Linearly parameterized bandits.
\newblock {\em Mathematics of Operations Research}, 2010.

\bibitem{rusmevichientong2006adaptive}
P.~Rusmevichientong and D.~P. Williamson.
\newblock An adaptive algorithm for selecting profitable keywords for
  search-based advertising services.
\newblock In {\em Proceedings of the ACM EC 2006}.

\bibitem{turney2000learning}
P.~D. Turney.
\newblock Learning algorithms for keyphrase extraction.
\newblock {\em Information Retrieval}, 2000.

\bibitem{witten1999KEA}
I.~Witten, G.~Paynter, E.~Frank, C.~Gutwin, and C.~Nevill-Manning.
\newblock Kea: Practical automatic keyphrase extraction.
\newblock In {\em Proceedings of the ACM DL 1999}.

\bibitem{wu2009advertising}
H.~Wu, G.~Qiu, X.~He, Y.~Shi, M.~Qu, J.~Shen, J.~Bu, and C.~Chen.
\newblock Advertising keyword generation using active learning.
\newblock In {\em Proceedings of the ACM WWW 2009}.

\bibitem{yih2006finding}
W.-t. Yih, J.~Goodman, and V.~R. Carvalho.
\newblock Finding advertising keywords on web pages.
\newblock In {\em Proceedings of the ACM WWW 2006}.

\bibitem{yuan2012sequential}
S.~Yuan and J.~Wang.
\newblock Sequential selection of correlated ads by pomdps.
\newblock In {\em Proceedings of the ACM CIKM 2012}.

\bibitem{zaragoza2004microsoft}
H.~Zaragoza, N.~Craswell, M.~Taylor, S.~Saria, and S.~Robertson.
\newblock Microsoft cambridge at trec-13: Web and hard tracks.
\newblock In {\em Proceedings of the TREC 2004}. Citeseer, 2004.

\end{thebibliography}
}

\end{document}